\documentclass[letterpaper, 12pt]{article}
\usepackage[margin=1in]{geometry}

\usepackage{times}
\usepackage{helvet}
\usepackage{courier}
\usepackage[hyphens]{url}
\usepackage{graphicx}
\usepackage{natbib}
\usepackage{caption}
\usepackage{booktabs}
\usepackage{multirow}
\usepackage{xcolor}

\title{Institutional Floors and Partisan Lenses: Cross-National Online Discourse\\
on Political Violence in France and the United States}

\author{Andrew Yen Chang\\
Department of Data Science, University of California, Berkeley\\
\texttt{andrewchang382@berkeley.edu}}

\date{}

\begin{document}

\maketitle

\begin{abstract}
This paper studies how online discussion shapes and assesses political violence across different settings, particularly how moral evaluation, as a social perception, varies across institutional contexts \cite{fiske1993}. We take France and the United States as case studies, both democracies, and three incidents of political violence: the 2020 killing of Samuel Paty in France, the 2025 shooting of Charlie Kirk in the United States, and the 2026 murder of Quentin Deranque in France \cite{marshall2024,grant2025,ouestfrance2026}. Using publicly available posts on Instagram and Facebook, we use GPT-4o-mini for zero-shot classification and social network analysis \cite{meta2024,openai2024}. Our research demonstrates clear cross-national differences in how moral values are perceived, the emotional intensity expressed, the framing of institutions, and the structure of semantic networks. In France, the discourse tends to focus on the victim's civic role rather than their political affiliation, whilst in the U.S., the conversation is more ideologically divided, with moral judgments frequently reflecting partisan lines. By comparing the two French cases---a civic victim (Paty) versus the politically-affiliated victim (Deranque)---we find evidence consistent with the \textit{civic floor hypothesis}, which demonstrates France's institutional framework upholds a cross-partisan civic baseline regardless of the victim's political ties. We conclude by analyzing the implications of computational social perception for multilingual NLP and by exploring moral judgment in cross-national digital political discourse.
\end{abstract}

%-----------------------------------------------------------------------
\section{Introduction}
%-----------------------------------------------------------------------

Political violence stirs up many sentiments that reflect institutional values and norms, culture, and social identity \cite{fiske1993,graham2013}. The frames for which the public sphere evaluates such events reveal the normative structural makeup and mindsets of their societies \cite{habermas1974,fiske1993}.

Online discourse provides a window into this phenomenon. Social media platforms serve as new public spheres for moral judgment, emotional responses, and blame attribution. By analyzing how public actors respond to instances of political violence, we can investigate how differing democratic value structures mold cultural interpretation at scale.

An important distinction aids such investigation: the violent event and its interpreted meaning. The event is an objective occurrence, while its perceived meaning---as a civic atrocity, a partisanship issue, religious extremism, or a tragic accident---is not simply determined by the event's occurrence, but rather by the perception and interpretation that observers construct through their respective cultural and institutional structures \cite{fiske1993}. These frameworks themselves can be viewed as lenses colored by culture and institutions \cite{fiske1993}. This paper studies this specific perception layer, treating evaluation and framing as forms of social perception that depend on societal norms, and that in turn inform us about the societies from which they emerge \cite{fiske1993}.

Using three cases of political violence across two national contexts---France and the United States---we examine the 2020 beheading of French schoolteacher Samuel Paty, the 2025 shooting of American conservative activist Charlie Kirk, and the 2026 murder of French far-right activist Quentin Deranque \cite{gordon2001,marshall2024,grant2025,ouestfrance2026}. The former two are used in cross-national comparison, whilst the latter serves as an empirical test of whether observed differences reflect national value structures rather than merely the victim's type.

We pose three research questions. \textbf{RQ1:} How do public discussions of political violence differ between France and the United States in terms of dominant frames? \textbf{RQ2:} How do these frames reflect their respective political and cultural models? \textbf{RQ3:} What do differences in discourse composition reveal about how political violence is perceived and debated in each public sphere?

Our main argument is that the meaning of political violence is constructed through pre-existing institutional and cultural values and norms. French discourse exhibits a civic-republican structure that fosters strong moral consensus regardless of the type of victim. American discourse displays a divided partisanship structure that leads to evaluative fragmentation. These differences are observable in online public posts.

%-----------------------------------------------------------------------
\section{Theoretical Background}
%-----------------------------------------------------------------------

\subsection{French and American Democratic Value Structures}

France and the U.S. have similar yet distinct models of democratic institutions and values \cite{gordon2001}. This creates a stable environment to evaluate their respective differences in response to acts of political violence. The French Fifth Republic, under the motto \textit{libert\'e, \'egalit\'e, fraternit\'e}, is organized as a unitary state representing its people through Universal Republicanism, wherein citizens are regarded as equal individuals while group identities receive less institutional recognition \cite{gcp2018}. \textit{La\"{i}cit\'e}---the French concept of secularism that removes religion from public institutions---is unique to France compared to the United States, where religion is tolerated in public life \cite{zoller2006}. France also enforces Race-Blind governance, with the collection of racial data strictly restricted \cite{labreck2021}. Together, these principles center political legitimacy on the Republic, its civic duties, and state authority \cite{laborde2008}.

The United States operates under a contrasting structure with extensive Freedom of Speech and Expression, Freedom of Religion in public life, and a pluralist identity model that allows racial, religious, and cultural identities as legitimate bases for political mobilization \cite{devos2014}. The decentralized federal system allows for great contrasts in governmental views, and political legitimacy may center on party partisanship, expressive freedoms, and divided group-based identities rather than a unified civic universalism \cite{nivola2005}.

\subsection{The Republican State as Institutional Embodiment of Civic Value}

A theoretical premise of our institutional floor hypothesis is French republican theory, which holds that the state and civic values are not separate domains \cite{daly2015}. The Republic is not merely a form of government, but an institution embodying the general will and civic universalism of the French people \cite{daly2015}. French rights discourse is classically articulated through republican constitutionalism and the Declaration, rather than primarily through the U.S.-style bill-of-rights framing. Rights are defined by the Declaration of the Rights of Man and of the Citizen and the Constitution of the Fifth Republic, and upheld by the state \cite{daly2015}. \textit{La\"{i}cit\'e} is not a private preference but a constitutional obligation enforced by the state \cite{zoller2006}. Civic education is similarly administered by the state as a republican duty \cite{daly2015}.

Such a tradition carries direct empirical implications. When a violent event occurs in France, the republican response is expressed in French civic language---\textit{la\"{i}cit\'e}, republican values, civic duty---or in state authority language through discourse on governmental response, institutional legitimacy, and state protection. Both are expressions of the French Republic. We therefore conceptualize \textit{institutional activation} as discourse that invokes either civic universalist norms or state authority as the guardian of those norms. This framing is rooted in French Republican Theory and is not constructed post hoc from the data. The Deranque case serves as a stress test of whether this dual notion of civic and institutional response is elevated for a partisan victim relative to a civic victim such as Paty.

\subsection{Social Media as a Public Sphere of Moral Judgment}

Social media platforms can signal and amplify moral judgment, emotional responses, and blame attribution following acts of political violence, making these values now quantifiable at scale \cite{habermas1974}. We integrate computational frame detection and comparative political theory into a unified methodology for studying social perception across national contexts.

\subsection{Moral Foundations and Evaluative Framing}

Moral Foundation Theory notes that care/harm, fairness, loyalty, authority, and sanctity are universal moral dimensions but vary in interpretation across cultures \cite{graham2013}. We extend this concept by asking how national institutional structures shape the moral framing that dominates public discourse for acts of political violence, whilst varying the national institutional contexts and, for the two French cases, the victim type.

\subsection{Moral Evaluation as Social Perception}

We situate our study within the social perception framework, where moral evaluation is a paradigmatic case of social perception \cite{fiske1993}. When assessing an act of political violence, observers apply a normative scheme shaped by cultural membership and institutional socialization---not a fully objective response to the death of a person. This creates differing perceptual frames that may vary across interpretive communities, in this case the French and the Americans. By comparing frame distributions across national contexts and victim types, we investigate how social perceptions may vary as a function of institutional values and design.

%-----------------------------------------------------------------------
\section{Comparative Cases}
%-----------------------------------------------------------------------

\subsection*{Samuel Paty, France (2020)}

On October 16th, 2020, Samuel Paty, a middle school teacher, was beheaded outside his school for using caricatures of the Prophet Muhammad in a lecture on freedom of expression during a civics class \cite{france242024}. His murder was interpreted widely as an attack on the French Republic---on \textit{la\"{i}cit\'e}, republican education, and state authority---triggering national mourning and intense public debate \cite{breeden2020}. Paty thus represents a \textit{civic victim}, killed in the line of his institutional function, with his death directly implicating republican values.

\subsection*{Charlie Kirk, United States (2025)}

Charlie Kirk, the founder of Turning Point USA, was shot and killed during a political rally on September 10, 2025 \cite{ax2025}. Associated with the Make America Great Again movement and Christian nationalism, his murder was widely framed through political partisanship, polarization, and ideological conflict \cite{bbcnews2025}. Kirk therefore represents a \textit{partisan victim}, a figure whose public identity was surrounded by political and ideological contestation.

\subsection*{Quentin Deranque, France (2026)}

A 23-year-old far-right activist, Quentin Deranque, died in Lyon on February 14, 2026, after sustaining injuries during a clash between far-right and left-wing groups \cite{ouestfrance2026}. His death triggered national political reactions at the highest levels of France, including a minute of silence in the National Assembly and President Macron calling on all parties to condemn political violence \cite{ouestfrance2026}. Deranque represents a theoretically critical test case: unlike Paty, he was a partisan figure killed during ideological conflict, providing a direct empirical test of the institutional floor hypothesis.

A potential objection is that Paty was a civic victim whilst Kirk and Deranque were partisan, suggesting that observed differences may reflect victim type rather than national institutional contexts. We address this on two grounds. First, victim type is not fully independent of national institutional context: the category of civic death is more institutionally valorized in France than in the United States. Second, our claim is not that civic victims produce civic discourses, but rather that French institutional architecture maintains a trans-partisan register robust to victim-type-specific framing. If French discourse around a far-right partisan victim still exhibits meaningful institutional framing, that constitutes evidence for the institutional floor independent of victim type.

%-----------------------------------------------------------------------
\section{Methods}
%-----------------------------------------------------------------------

\subsection{Data Collection}

We collected public posts from Instagram and Facebook via Meta's Content Library API \cite{meta2024}, with collection windows running from 3 days before each event through 31 days after: October 13--November 16, 2020 for Paty; September 7--October 11, 2025 for Kirk; and February 11--March 16, 2026 for Deranque. Posts were filtered by language (French for Paty and Deranque, English for Kirk). We sampled 1,000 posts per platform per case ($n = 2{,}000$ per case; $n = 6{,}000$ for the full matched analysis) with a fixed random seed of 42. Platform files were maintained separately for robustness checks.

\subsection{LLM Zero-Shot Classification}

We utilized GPT-4o-mini for zero-shot classification \cite{openai2024}, systematically coding three dimensions for each post: (1) the dominant \textit{frame}, drawn from seven theoretically-defined categories; (2) the dominant \textit{emotion}, using six categories; and (3) \textit{moral evaluation}, divided into four categories. The model temperature was set to 0 to ensure reproducibility, and all prompts are provided in Appendix~\ref{app:prompt}.

Drawing on the theoretical discussion above, we combine the Civic and State frame categories into a composite \textit{institutional register}, reflecting their non-distinct domain within French Republican tradition. The full frame taxonomy is as follows:

\begin{itemize}
    \item \textbf{Political Violence} --- discussion of the act as an instance of political violence
    \item \textbf{Promoting Violence} --- content endorsing or celebrating the violence
    \item \textbf{Party Partisanship} --- framing through partisan political identification
    \item \textbf{State} --- framing through state authority, government response, or institutional legitimacy
    \item \textbf{Nationalism} --- framing through national identity or patriotism
    \item \textbf{Religion} --- framing through religious identity or doctrine
    \item \textbf{Civics} --- framing through civic values, rights, education, or democratic norms
\end{itemize}

Emotion categories: anger, fear/anxiety, sadness/grief, moral outrage/disgust, contempt/mockery, neutral/unclear. Moral evaluation categories: condemn, justify, ambivalent, unclear.

\subsection{Validation Cross-Check}

To validate GPT-4o-mini outputs, we manually coded 100 posts per language corpus (Paty for French; Kirk for English) and computed Cohen's kappa against the LLM labels \cite{cohen1960}. Given identical prompting and model architecture across all three cases, validation on one corpus per language is sufficient to establish reliability across the full dataset. Kappa scores are reported in the Results section.

Our analysis proceeded in three steps. First, we conducted a distributional comparison via Chi-Square tests to assess statistical significance across all pairwise case comparisons. Second, we performed theoretical interpretation with particular attention to the Paty--Deranque comparison as a test of the institutional floor hypothesis. Third, we conducted human-coder vs.\ GPT-4o-mini validation to assess classification reliability. This allows us to address the three RQs. Platform distributions were checked for robustness in Appendix~\ref{app:platform}.

%-----------------------------------------------------------------------
\section{Results}
%-----------------------------------------------------------------------

\subsection{Moral Evaluation}

\begin{table}[htbp]
\centering
\caption{Moral evaluation distributions by case (\%).}
\label{tab:moral}
\begin{tabular}{lrrr}
\toprule
\textbf{Evaluation} & \textbf{Paty} & \textbf{Kirk} & \textbf{Deranque} \\
\midrule
Condemn    & 92.4 & 65.2 & 77.3 \\
Justify    &  1.5 &  4.2 &  8.1 \\
Ambivalent &  1.0 &  3.0 &  4.8 \\
Unclear    &  5.1 & 27.6 &  9.8 \\
\bottomrule
\end{tabular}
\end{table}

Table~\ref{tab:moral} reports the distributions of moral evaluations for each case. All pairwise comparisons reached statistical significance (Paty vs.\ Kirk: $\chi^2=727.2$, $df=3$, $p<.001$; Paty vs.\ Deranque: $\chi^2=334.0$, $df=3$, $p<.001$; Deranque vs.\ Kirk: $\chi^2=85.9$, $df=3$, $p<.001$).

The Paty corpus displays near-unanimous condemnation (92.4\%), with only 5.1\% unclear responses, indicating minimal evaluative ambiguity. The Kirk corpus shows a notably lower condemnation rate (65.2\%) and a markedly higher proportion of unclear or ambiguous evaluations (27.6\%), reflecting significant fragmentation in American discourse. The Deranque corpus falls between these two cases, with condemnation at 77.3\%---appreciably higher than Kirk despite Deranque also being a partisan victim---providing initial support for the institutional floor hypothesis.

\begin{figure}[t]
\centering
\includegraphics[width=0.95\columnwidth]{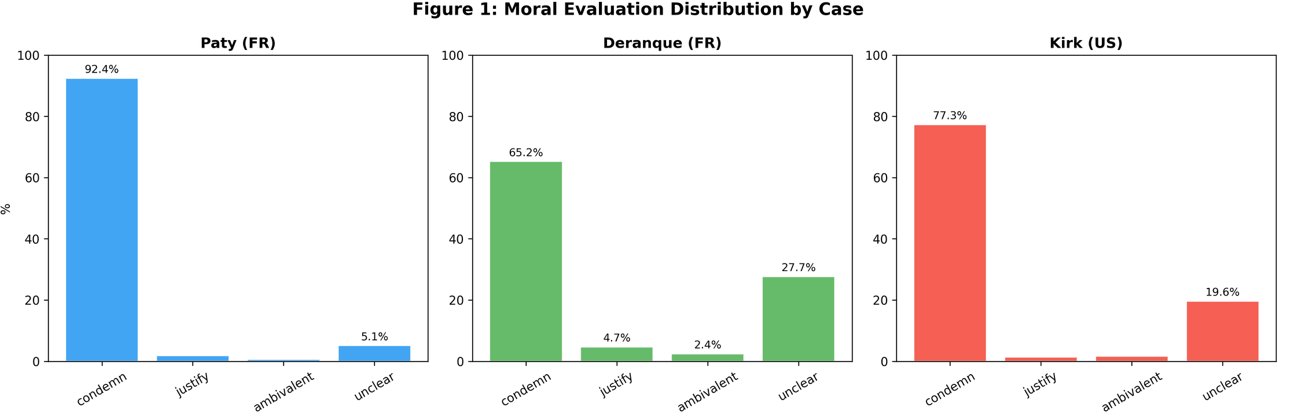}
\caption{Moral evaluation distributions across the three cases. Paty (FR) shows near-unanimous condemnation, while Kirk (US) displays substantially higher evaluative ambiguity. Deranque (FR) occupies an intermediate position, consistent with the institutional floor hypothesis.}
\label{fig:moral}
\end{figure}

\subsection{Emotional Evaluation}

\begin{table}[htbp]
\centering
\caption{Emotional evaluation distributions by case (\%).}
\label{tab:emotion}
\begin{tabular}{lrrr}
\toprule
\textbf{Emotion} & \textbf{Paty} & \textbf{Kirk} & \textbf{Deranque} \\
\midrule
Sadness/Grief         & 43.1 & 18.4 & 26.6 \\
Moral Outrage/Disgust & 40.7 & 32.6 & 41.2 \\
Anger                 &  8.2 & 12.4 & 18.5 \\
Fear/Anxiety          &  3.1 & 14.1 &  6.3 \\
Contempt/Mockery      &  1.8 &  6.2 &  3.6 \\
Neutral/Unclear       &  3.1 & 16.3 &  3.8 \\
\bottomrule
\end{tabular}
\end{table}

Table~\ref{tab:emotion} shows the emotional distributions. All pairwise comparisons were significant (Paty vs.\ Kirk: $\chi^2=522.7$, $df=5$, $p<.001$; Paty vs.\ Deranque: $\chi^2=325.0$, $df=5$, $p<.001$; Deranque vs.\ Kirk: $\chi^2=201.6$, $df=5$, $p<.001$).

Both French cases are characterized by grief and moral outrage. Paty shows sadness/grief (43.1\%) and moral outrage (40.7\%), and Deranque shows moral outrage (41.2\%) and sadness/grief (26.6\%). American discourse exhibited higher fear/anxiety (14.1\%) and neutral/unclear responses (16.3\%), consistent with evaluative fragmentation. The similarity between the two French cases despite differing victim types suggests that national institutional context shapes the emotional register more than victim characteristics.

\begin{figure}[t]
\centering
\includegraphics[width=0.95\columnwidth]{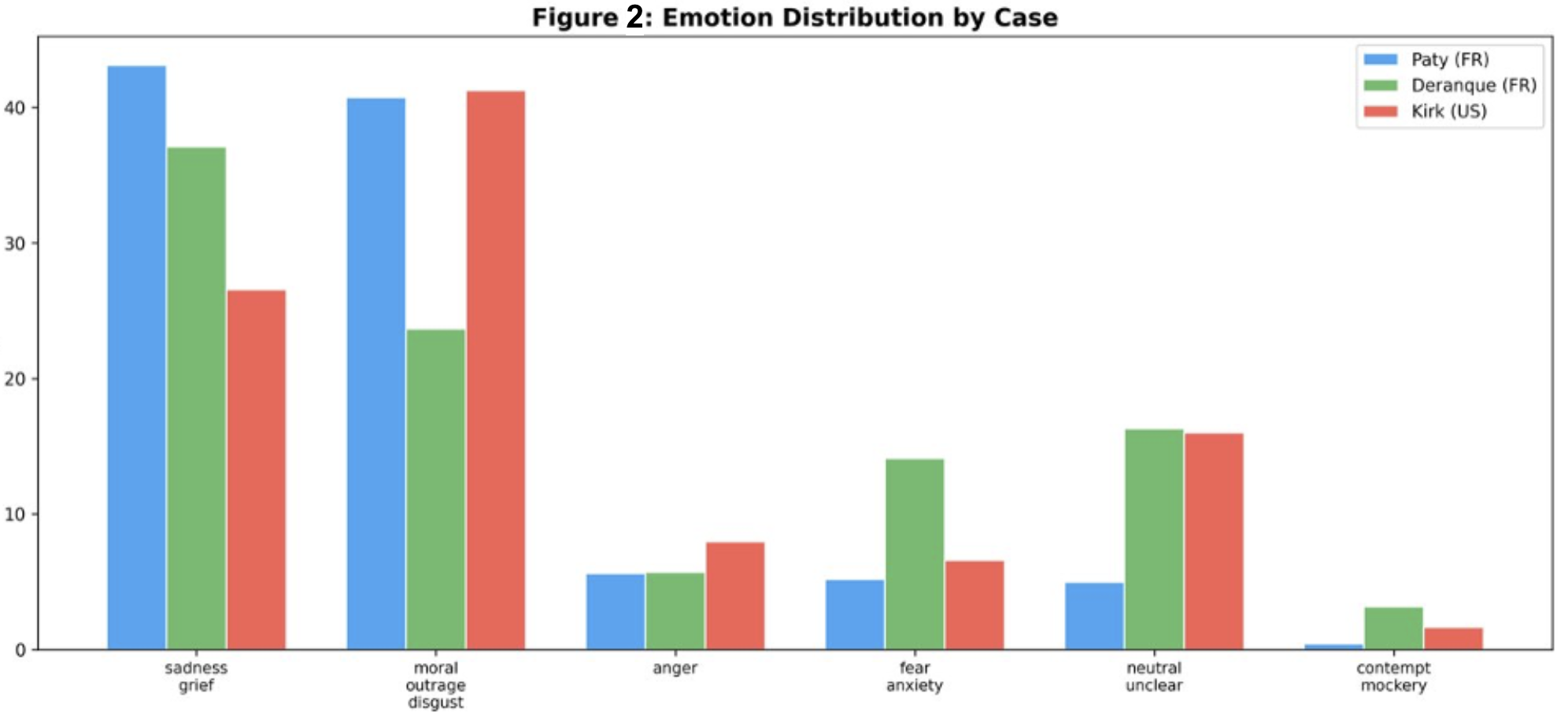}
\caption{Emotion distributions across all three cases. Both French cases (Paty and Deranque) are dominated by sadness/grief and moral outrage, while the American case (Kirk) shows markedly higher fear/anxiety and neutral/unclear responses, reflecting evaluative fragmentation.}
\label{fig:emotion}
\end{figure}

\subsection{Frame Evaluation and the Institutional Floor}

\begin{table}[htbp]
\centering
\caption{Frame distributions by case (\%).}
\label{tab:frames}
\begin{tabular}{lrrr}
\toprule
\textbf{Frame} & \textbf{Paty} & \textbf{Kirk} & \textbf{Deranque} \\
\midrule
Political Violence  & 23.1 & 46.8 & 38.2 \\
Civics              & 39.0 &  5.2 &  5.2 \\
State               & 18.6 &  9.8 & 20.4 \\
Party Partisanship  &  4.2 & 17.4 & 14.8 \\
Nationalism         &  8.3 &  7.4 & 12.6 \\
Religion            &  5.1 &  8.6 &  4.6 \\
Promoting Violence  &  1.7 &  4.8 &  4.2 \\
\midrule
\textit{Institutional (Civic+State)} & \textit{57.6} & \textit{15.0} & \textit{25.6} \\
\bottomrule
\end{tabular}
\end{table}

Table~\ref{tab:frames} illustrates the frame distributions. All pairwise comparisons were significant (Paty vs.\ Kirk: $\chi^2=1492.2$, $df=6$, $p<.001$; Paty vs.\ Deranque: $\chi^2=1749.3$, $df=7$, $p<.001$; Deranque vs.\ Kirk: $\chi^2=303.8$, $df=7$, $p<.001$).

The Paty corpus is dominated by Civics (39.0\%), Political Violence (23.1\%), and State (18.6\%). The Kirk corpus is dominated by Political Violence (46.8\%) and Party Partisanship (17.4\%), with Civics at only 5.2\%.

The Deranque corpus reveals the nuance of the institutional floor hypothesis. Explicit Civic framing collapsed to 5.2\%---identical to Kirk---but State framing remained high at 20.4\%, comparable to and even slightly exceeding Paty (18.6\%), and substantially higher than Kirk (9.8\%). Combining Civic and State framing into a composite institutional register, France maintains 57.6\% for Paty and 25.6\% for Deranque, compared to 15.0\% for Kirk. This pattern suggests that the French institutional floor operates through two mechanisms: civic values for civic victims and state authority for partisan victims. In the French republican tradition, these are theoretically linked---the state is the institutional embodiment of civic values---making both mechanisms part of the same underlying institutional architecture \cite{daly2015}.

\begin{figure*}[t]
\centering
\includegraphics[width=0.95\textwidth]{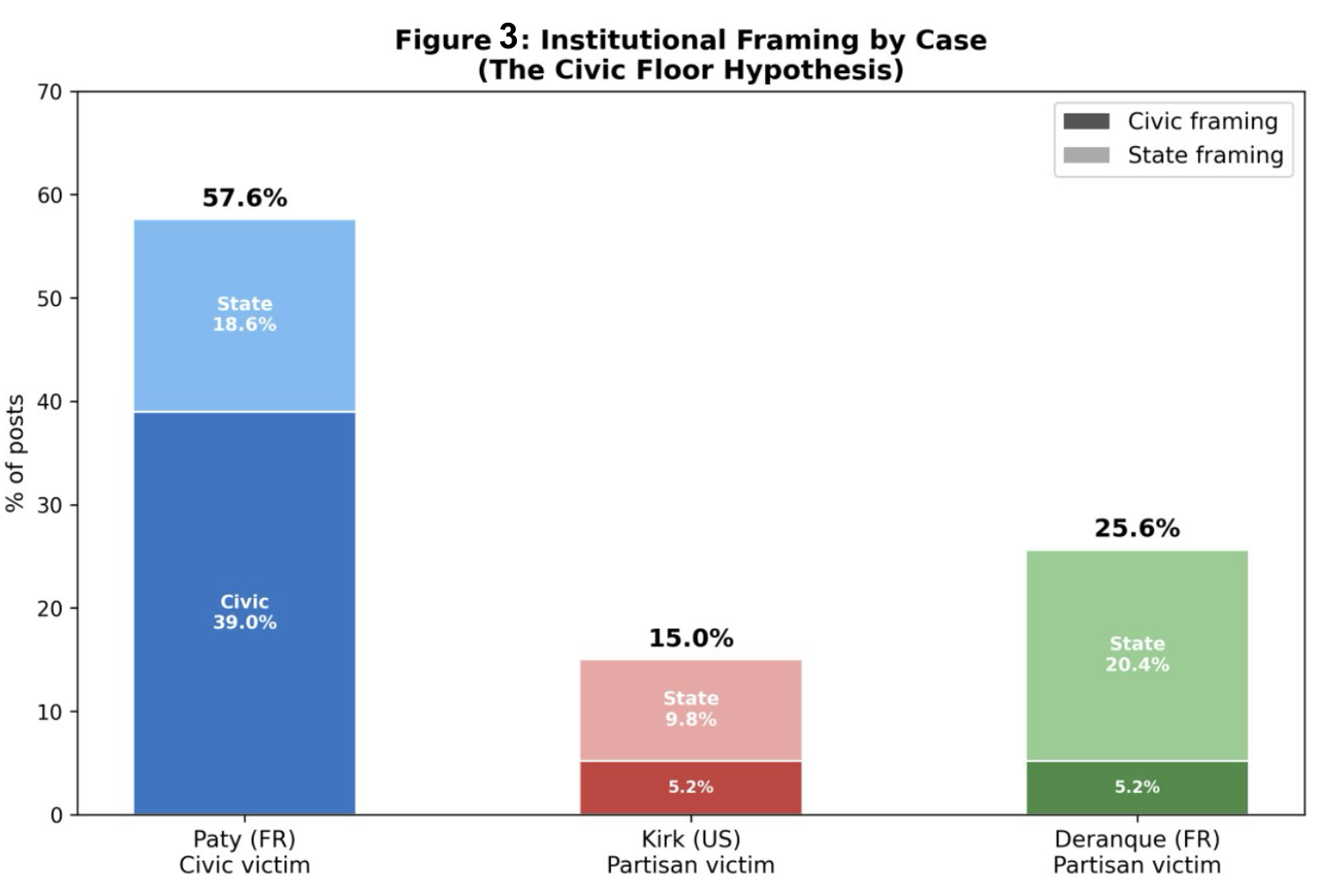}
\caption{Composite institutional framing (Civic + State) by case. Paty (FR) reaches 57.6\%, Deranque (FR) 25.6\%, and Kirk (US) only 15.0\%. The shift from civic to state framing between the two French cases reflects the dual mechanism of the institutional floor: civic language for civic victims, state authority language for partisan victims.}
\label{fig:institutional}
\end{figure*}

\subsection{Validation}

We verified GPT-4o-mini reliability by manually coding 100 language-filtered posts per corpus and computing Cohen's kappa \cite{cohen1960}; full results are reported in Appendix~\ref{app:validation}. For the French corpus (Paty): frame agreement was moderate ($\kappa=0.445$, 64\% accuracy), emotion agreement was moderate ($\kappa=0.528$, 71\%), and moral evaluation agreement was fair ($\kappa=0.065$, 34\%). For the English corpus (Kirk): frame agreement was fair ($\kappa=0.387$, 56\%), emotion agreement was fair ($\kappa=0.360$, 50\%), and moral evaluation agreement was fair ($\kappa=0.150$, 50\%).

The consistently lower moral evaluation kappa may reflect that human coders applied the \textit{unclear} label more liberally, while the LLM showed greater commitment to \textit{condemn}. Frame and emotion agreement were notably higher in the French corpus, perhaps reflecting the greater semantic coherence of Paty's discourse compared to the fragmented Kirk corpus. The moderate frame and emotion kappas for French support the reliability of our civic and institutional framing results, which are central to our argument.

Interestingly, the human coder foregrounded the violent act itself in frame classification, while the LLM foregrounded the frames implicated by the political violence. This difference is consistent with our theoretical claim that frame classification reflects socially conditioned perceptions rather than objective event description \cite{fiske1993}.

\subsection{Lexicon-Based Robustness Check: NRC, eMFD, and VADER}

To see whether LLM zero-shot emotion classifications capture nuance beyond lexical word associations, we utilized three lexicon-based methods: the NRC Word-Emotion Association Lexicon \cite{mohammad2013}, the extended Moral Foundations Dictionary \cite{hopp2021}, and VADER sentiment analysis \cite{hutto2014}. The NRC method gives binary word-level associations with eight basic emotions. We used the English NRC lexicon for the Kirk corpus and the NRC French translation for the Paty and Deranque corpora. eMFD and VADER are English-only and were applied to the Kirk corpus only; full results are in Appendix~\ref{app:lexicon}.

Our results for the NRC analysis differ greatly from the LLM classification because it is unable to take context into account when forming its judgment. For example, trust dominated the NRC emotion for both French corpora---20.8\% for Paty and 22.7\% for Deranque respectively. For Kirk, fear and anger dominate at 18.2\% and 15.2\% respectively. By contrast, LLM classification identified sadness/grief and moral outrage as the dominant emotions across all corpora, with trust not a classification category. NRC also severely undercounted sadness (Paty 6.6\%, Kirk 11.2\%, Deranque 10.0\%) and disgust (Paty 6.2\%, Kirk 3.6\%, Deranque 9.5\%) relative to LLM scores of 43.1\%/37.1\%/26.6\% for sadness and 40.7\%/23.6\%/41.2\% for moral outrage.

To investigate why trust is so dominant, we examined the highest-frequency trust words in the French corpora. For Paty, these were: \textit{professeur} (873), \textit{libert\'e} (684), \textit{foi} (294), \textit{police} (290), \textit{pr\'esident} (257), \textit{\'ecole} (195), \textit{proph\`ete} (137), \textit{honneur} (119), and \textit{nation} (109). For Deranque, similar institutional and civic vocabulary drives the trust score. None of these words express trust as a felt emotion. Rather, they are the constitutive vocabulary of the civic violent event: a teacher killed in the line of his institutional function, invoking the French values of freedom of speech, state authority, and civic discourse. NRC classifies them as trust because in general usage words like ``professor'', ``freedom'', ``president'', and ``honour'' correlate with trust---but in the context of political violence, they carry grief, outrage, and civic invocation.

Given the limitations of lexicon-based classifiers, LLM zero-shot processing remains the primary instrument for this study, as it requires contextual knowledge to classify meaning correctly. This also yields a methodological finding: NRC-based emotion identification can systematically misclassify the emotional register of political violence discourse by conflating civic and institutional vocabulary with trust.

Notably, NRC and LLM converge on one finding: the Kirk corpus shows substantially higher anger (NRC: 15.2\% vs.\ 8.2\% and 12.8\% for the French cases) and fear (NRC: 18.2\% vs.\ 8.5\% and 12.7\%) than both French corpora, providing cross-method support for the emotional fragmentation finding. eMFD moral foundation scores for the Kirk corpus are reported in Appendix~\ref{app:lexicon}; overall coverage was sparse (689 words) and scores were too low to support substantive interpretation.

%-----------------------------------------------------------------------
\section{Discussion}
%-----------------------------------------------------------------------

\subsection{The Institutional Floor Hypothesis}

Our central contribution is the \textit{institutional floor hypothesis}: French republican structure maintains a trans-partisan institutional framing register for political violence discourse, one that is robust to victim type though sensitive in its expression. Both French cases yielded higher condemnation (92.4\% and 77.3\%) and combined institutional framing (57.6\% and 25.6\%) than the Kirk corpus (65.2\% condemnation and 15.0\% institutional framing).

The type of institutional framing shifts between the two French cases. For Paty, the floor is expressed via explicit civic language (Civics 39.0\%). For Deranque, explicit civic framing falls (5.2\%), but state authority framing compensates (20.4\%). This shift is theoretically coherent: in the French republican tradition, state authority and civic values are not fully distinct \cite{daly2015}. Thus, when a victim dies a civic death, French discourse invokes civic values directly; when the victim is a partisan figure, French discourse invokes the state as guarantor of those values. Both nonetheless represent the republican institutional response to political violence.

The American case shows no equivalent mechanism. Kirk's shooting produced high partisan framing (17.4\%) but low institutional framing (15.0\%), consistent with a discourse environment lacking a trans-partisan institutional register. We caution, however, that this comparison rests on a single American case; a civic American case would be required to fully determine whether the absence of an institutional floor is structural or event-specific.

\subsection{Diverging Moral Consensus}

French discourse converges on strong condemnation across both cases, while American discourse fragments into evasive ambiguity (27.6\% unclear). This pattern is consistent with differences in interpretive lenses across national value systems. Institutional framing---whether civic or state---appears to function as a consensus-generating mechanism. Partisan lenses, by contrast, may sustain competing interpretive frames that preclude convergence.

Our framework generated a falsifiable prediction: a highly partisan French public figure subjected to political violence should still exhibit a meaningfully higher institutional framing floor and condemnation rate than any comparable American case. The Deranque case provides partial support; a more prominent partisan victim would constitute a stronger test.

\subsection{Institutional Framing and Political Legitimacy}

The prominence of institutional framing in French discourse and partisan framing in American discourse is consistent with long-standing differences in the organization of political legitimacy. In France, republican legitimacy has historically been anchored in centralized state authority and civic universalism, making attacks more likely to be interpreted as violations of a shared civic order \cite{daly2015}. In the United States, political legitimacy is more often organized through pluralist and competitive partisan conflict, making political violence easier to process as part of an ongoing political struggle rather than a universally condemned rupture \cite{almond1963}. This finding is consistent with democratic theory that emphasizes the role of civic culture in maintaining shared civic baselines \cite{almond1963}.

\subsection{Implications for Computational Social Perception}

Our findings argue for treating frame and emotional distributions as \textit{perceptual distributions} rather than as event descriptions. Differences in the events themselves cannot explain the observed distributions, as all three involve the deaths of prominent figures in democratic contexts. Rather, differing perceptual frameworks that observers possess in different national contexts may be explanatory \cite{fiske1993}.

This reframing has implications for cross-cultural NLP, where tasks such as stance detection, sentiment analysis, and frame classification are often treated as having specific ``correct'' labels. Our results suggest that, for politically charged content, disagreements between annotators from different national contexts may reflect genuine perceptual differences rooted in differing norms, rather than annotation error. Attempting to resolve such variances may erase the most theoretically significant signal in the data. Our NRC robustness check extends this argument further: even established lexicon-based tools can import cultural assumptions that distort cross-lingual analysis, suggesting LLM zero-shot classification may be better suited for political violence discourse across languages.

%-----------------------------------------------------------------------
\section{Limitations}
%-----------------------------------------------------------------------

Several limitations should be noted. First, our data consists of public posts from Meta's Content Library, which may not be representative of general public opinion. Second, the three cases span 2020 to 2026, during which platform dynamics, moderation, and algorithmic changes may contribute to observed differences unrelated to institutional structure. Third, the cross-national comparison rests on a single American case without within-US variation; a complete 2$\times$2 design would require a civic American victim case, which we leave for future work. Fourth, GPT-4o-mini validation was conducted on 100 posts per language corpus, and individual classifications may contain errors due to ambiguity or ironic content. Fifth, the cases differ in national salience---Paty was a national trauma, Kirk was a partisan rally shooting, and Deranque was a street clash---and these differences may also account for some observed variation. Sixth, the NRC French lexicon is a machine translation of the English original and has not been independently validated for French political discourse; eMFD and VADER are English-only and were therefore restricted to the Kirk corpus. Finally, as with all LLM-based annotation, exact replication requires access to the same model version; OpenAI does not guarantee identical outputs across model updates, though temperature was set to 0 to minimize within-version variance.

%-----------------------------------------------------------------------
\section{Conclusion}
%-----------------------------------------------------------------------

This paper examined how differing democratic value structures shape online moral evaluation and framing of political violence across three cases in two national contexts. We find systematic differences in condemnation, emotional, and frame distributions consistent with theoretical accounts of national institutional value systems.

Our principal contribution is the articulation and empirical validation of the \textit{institutional floor hypothesis}: French republican infrastructure supports a trans-partisan institutional framing mechanism robust to victim type. This mechanism manifests through the consistent activation of civic values for civic victims and the mobilization of state authority for partisan victims, producing markedly higher institutional framing and moral consensus than observed in American discourse. By demonstrating this pattern across both French cases, we advance understanding of how national institutional contexts shape the normative and emotional contours of online responses to political violence.

These findings contribute to the comparative sociology of moral evaluation, the computational study of social perception, the study of political violence, and cross-cultural NLP methodology. Framing and emotions in online discourse are perceptual distributions reflecting how socially conditioned observers make meaning of public events---not objective event descriptions. We hope this encourages further work at the intersection of democratic theory, social perception, and computational social science.

%-----------------------------------------------------------------------
\bibliographystyle{plain}
\bibliography{references}

%-----------------------------------------------------------------------
\appendix
%-----------------------------------------------------------------------

\section{GPT-4o-mini Classification Prompt}
\label{app:prompt}

The following prompt was used for all zero-shot classifications across all three corpora. Temperature was set to 0 for full reproducibility.

\medskip
\noindent\textbf{System prompt:}
\begin{quote}
\texttt{You are a social science researcher coding social media posts about political violence. You will classify each post on three dimensions: frame, emotion, and moral evaluation. Always respond with valid JSON only. No explanation, no preamble, no markdown.}
\end{quote}

\noindent\textbf{Frame categories:} \texttt{political\_violence}, \texttt{promoting\_violence}, \texttt{party\_partisanship}, \texttt{state}, \texttt{nationalism}, \texttt{religion}, \texttt{civics}

\medskip
\noindent\textbf{Emotion categories:} \texttt{anger}, \texttt{fear\_anxiety}, \texttt{sadness\_grief}, \texttt{moral\_outrage\_disgust}, \texttt{contempt\_mockery}, \texttt{neutral\_unclear}

\medskip
\noindent\textbf{Moral evaluation categories:} \texttt{condemn}, \texttt{justify}, \texttt{ambivalent}, \texttt{unclear}

\medskip
\noindent\textbf{Response format:}
\begin{quote}
\texttt{\{"frame": "<label>", "emotion": "<label>", "moral\_evaluation": "<label>"\}}
\end{quote}

%-----------------------------------------------------------------------
\section{Platform Robustness Checks}
\label{app:platform}

Tables~\ref{tab:plat_frame}--\ref{tab:plat_moral} report frame, emotion, and moral evaluation distributions separately by platform (Facebook and Instagram) for each case. Distributions are broadly consistent across platforms within each case, supporting the robustness of the pooled results reported in the main text.

\begin{table*}[t]
\centering
\caption{Frame distributions by platform and case (\%).}
\label{tab:plat_frame}
\begin{tabular}{llrrr}
\toprule
\textbf{Platform} & \textbf{Frame} & \textbf{Paty} & \textbf{Kirk} & \textbf{Deranque} \\
\midrule
\multirow{8}{*}{Facebook}
 & Political Violence  & 23.4 & 60.8 & 40.1 \\
 & Civics              & 39.7 &  3.1 &  6.9 \\
 & State               & 19.4 &  9.8 & 23.2 \\
 & Party Partisanship  &  1.0 & 11.8 & 19.0 \\
 & Nationalism         &  3.6 &  2.5 &  5.2 \\
 & Religion            & 10.0 &  6.3 &  0.5 \\
 & Promoting Violence  &  1.9 &  3.4 &  0.4 \\
 & Unclear             &  1.0 &  2.3 &  4.7 \\
\midrule
\multirow{8}{*}{Instagram}
 & Political Violence  & 22.8 & 32.8 & 49.2 \\
 & Civics              & 38.2 &  7.3 &  3.4 \\
 & State               & 17.8 &  9.7 & 17.6 \\
 & Party Partisanship  &  0.5 & 23.1 & 23.0 \\
 & Nationalism         &  4.8 &  6.9 &  5.2 \\
 & Religion            & 14.2 & 13.7 &  0.4 \\
 & Promoting Violence  &  0.9 &  3.7 &  0.6 \\
 & Unclear             &  0.8 &  2.8 &  0.6 \\
\bottomrule
\end{tabular}
\end{table*}

\begin{table*}[t]
\centering
\caption{Emotion distributions by platform and case (\%).}
\label{tab:plat_emotion}
\begin{tabular}{llrrr}
\toprule
\textbf{Platform} & \textbf{Emotion} & \textbf{Paty} & \textbf{Kirk} & \textbf{Deranque} \\
\midrule
\multirow{6}{*}{Facebook}
 & Sadness/Grief         & 42.8 & 38.5 & 27.8 \\
 & Moral Outrage/Disgust & 38.3 & 22.6 & 35.5 \\
 & Anger                 &  5.6 &  4.3 &  6.5 \\
 & Fear/Anxiety          &  5.9 & 19.4 &  7.3 \\
 & Contempt/Mockery      &  0.7 &  2.7 &  1.7 \\
 & Neutral/Unclear       &  6.8 & 12.5 & 21.2 \\
\midrule
\multirow{6}{*}{Instagram}
 & Sadness/Grief         & 43.4 & 35.7 & 25.3 \\
 & Moral Outrage/Disgust & 43.2 & 24.7 & 47.0 \\
 & Anger                 &  5.6 &  7.1 &  9.4 \\
 & Fear/Anxiety          &  4.4 &  8.8 &  5.9 \\
 & Contempt/Mockery      &  0.2 &  3.6 &  1.6 \\
 & Neutral/Unclear       &  3.2 & 20.1 & 10.8 \\
\bottomrule
\end{tabular}
\end{table*}

\begin{table*}[t]
\centering
\caption{Moral evaluation distributions by platform and case (\%).}
\label{tab:plat_moral}
\begin{tabular}{llrrr}
\toprule
\textbf{Platform} & \textbf{Evaluation} & \textbf{Paty} & \textbf{Kirk} & \textbf{Deranque} \\
\midrule
\multirow{4}{*}{Facebook}
 & Condemn    & 89.5 & 69.8 & 72.1 \\
 & Justify    &  2.4 &  4.4 &  0.6 \\
 & Ambivalent &  0.8 &  2.8 &  1.4 \\
 & Unclear    &  7.4 & 23.0 & 25.9 \\
\midrule
\multirow{4}{*}{Instagram}
 & Condemn    & 95.3 & 60.7 & 82.5 \\
 & Justify    &  1.4 &  5.0 &  2.2 \\
 & Ambivalent &  0.4 &  2.0 &  2.0 \\
 & Unclear    &  2.8 & 32.3 & 13.3 \\
\bottomrule
\end{tabular}
\end{table*}

%-----------------------------------------------------------------------
\section{Validation Details}
\label{app:validation}

Table~\ref{tab:kappa} presents inter-rater reliability between the primary human coder and GPT-4o-mini on 100 language-filtered posts per corpus. Human coding was conducted independently after classification.

\begin{table}[t]
\centering
\caption{Human vs.\ GPT-4o-mini agreement ($n=100$ per corpus).}
\label{tab:kappa}
\small
\begin{tabular}{lrrlrrl}
\toprule
\textbf{Dim.} & \textbf{FR\%} & \textbf{FR$\kappa$} & \textbf{FR} & \textbf{EN\%} & \textbf{EN$\kappa$} & \textbf{EN} \\
\midrule
Frame  & 64\% & 0.445 & mod. & 56\% & 0.387 & fair \\
Emotion & 71\% & 0.528 & mod. & 50\% & 0.360 & fair \\
Moral  & 34\% & 0.065 & slight & 50\% & 0.150 & fair \\
\bottomrule
\end{tabular}
\end{table}

\noindent FR = French/Paty corpus; EN = English/Kirk corpus. Chance baselines: frame 14\% (7 categories), emotion 17\% (6 categories), moral evaluation 25\% (4 categories). Interpretation thresholds: $\kappa > 0.4$ moderate, $\kappa > 0.2$ fair, $\kappa > 0$ slight.

The lower moral evaluation kappa reflects the human coder's conservative use of the \textit{unclear} label where GPT-4o-mini committed to \textit{condemn}. The higher French agreement on frame and emotion plausibly reflects the greater semantic coherence of the Paty discourse relative to the more fragmented Kirk corpus, where posts ranged from breaking news to conspiracy theories to partisan commentary.

%-----------------------------------------------------------------------
\section{Lexicon-Based Supplementary Results}
\label{app:lexicon}

\begin{table}[htbp]
\centering
\caption{NRC emotion profile by case (\% of matched words).}
\label{tab:nrc}
\begin{tabular}{lrrr}
\toprule
\textbf{Emotion} & \textbf{Paty (FR)} & \textbf{Kirk (US)} & \textbf{Deranque (FR)} \\
\midrule
Trust        & 20.8 & 15.2 & 22.7 \\
Fear         &  8.5 & 18.2 & 12.7 \\
Anger        &  8.2 & 15.2 & 12.8 \\
Anticipation &  9.5 & 10.4 &  9.9 \\
Sadness      &  6.6 & 11.2 & 10.0 \\
Surprise     &  4.2 & 10.9 &  5.0 \\
Joy          &  5.3 &  5.4 &  3.6 \\
Disgust      &  6.2 &  3.6 &  9.5 \\
\bottomrule
\end{tabular}
\end{table}

\noindent\small French corpora scored using \texttt{French-NRC-EmoLex.txt} (machine-translated); Kirk scored using the English NRC lexicon. Trust inflation in French corpora is driven by event-specific civic vocabulary rather than felt trust emotion (see Section~5.5).

\medskip

\begin{figure*}[t]
\centering
\includegraphics[width=0.95\textwidth]{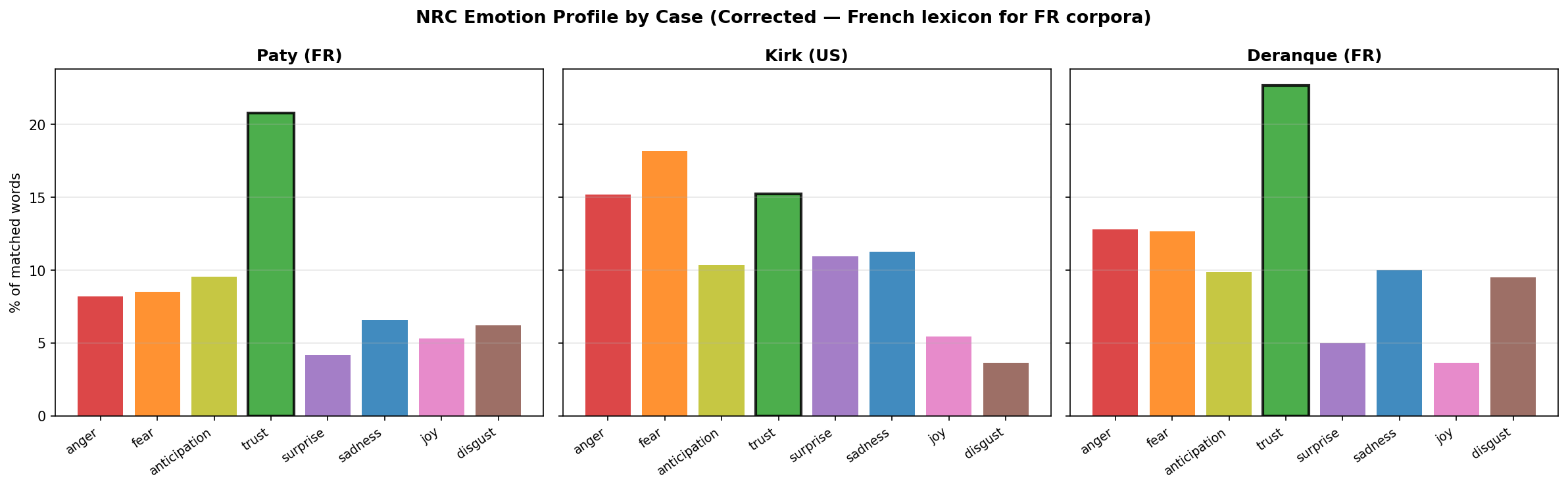}
\caption{NRC full emotion profile by case using language-appropriate lexicons (French NRC for FR corpora, English NRC for Kirk). Trust dominates both French corpora due to civic and institutional vocabulary --- \textit{professeur}, \textit{libert\'e}, \textit{police}, \textit{pr\'esident}, \textit{\'ecole} --- not genuine trust expression. Sadness and disgust/moral outrage, dominant in LLM classification, are severely undercounted.}
\label{fig:nrc}
\end{figure*}

\begin{table}[htbp]
\centering
\caption{eMFD moral foundation scores --- Kirk (US) only.}
\label{tab:emfd}
\begin{tabular}{lr}
\toprule
\textbf{Foundation} & \textbf{Mean Score} \\
\midrule
Care       & 0.150 \\
Fairness   & 0.099 \\
Sanctity   & 0.097 \\
Loyalty    & 0.088 \\
Authority  & 0.086 \\
\bottomrule
\end{tabular}
\end{table}

\noindent eMFD is English-only; French corpora excluded. Lexicon coverage: 689 words. Care dominance reflects the high volume of condolence and prayer posts in the Kirk corpus.

\begin{table}[htbp]
\centering
\caption{VADER sentiment analysis by case.}
\label{tab:vader}
\begin{tabular}{lrrr}
\toprule
\textbf{Case} & \textbf{\% Neg} & \textbf{\% Pos} & \textbf{\% Neu} \\
\midrule
Paty (FR)*     & 28.8 & 20.3 & 50.9 \\
Kirk (US)      & 55.6 & 26.8 & 17.5 \\
Deranque (FR)* & 42.0 & 16.6 & 41.4 \\
\bottomrule
\end{tabular}
\end{table}

\noindent Mean compound scores: Paty $-$0.080, Kirk $-$0.212, Deranque $-$0.172.

\noindent * French corpora: word-level neutrality $\approx$95\% indicates VADER cannot reliably score French text; post-level scores reported for completeness only. Kirk results are primary. Compound score $\geq$0.05 = positive, $\leq$$-$0.05 = negative, otherwise neutral.

\end{document}